\documentclass{amsart}

\usepackage[utf8]{inputenc}
\usepackage{amsmath, amsfonts, amsthm, amssymb}
\usepackage{tikz}
\usepackage{hyperref}
\usepackage{pst-all}
\usepackage{listings}

\title{On the Algebraic Properties of Flame Graphs}
\author{Gabriele N. Tornetta}
\date{February 2023}

\theoremstyle{definition}
\newtheorem{definition}{Definition}[section]
\newtheorem{example}{Example}[section]
\newtheorem{remark}{Remark}[section]

\newcommand{\supp}{\operatorname{supp}}
\newcommand{\diag}{\operatorname{diag}}

\begin{document}

\maketitle

\begin{abstract}
Flame graphs are a popular way of representing profiling data. In this paper we
propose a possible mathematical definition of flame graphs. In doing so, we gain
some interesting algebraic properties almost for free, which in turn allow us to
define some operations that can allow to perform an in-depth performance
regression analysis. The typical documented use of a flame graph is via its
graphical representation, whereby one scans the picture for the largest
plateaux. Whilst this method is effective at finding the main sources of
performance issues, it leaves quite a large amount of data potentially unused.
By combining a mathematical precise definition of flame graphs with some
statistical methods we show how to generalise this visual procedure and make the
best of the full set of collected profiling data.
\end{abstract}

\section{Introduction}

It is widely accepted that an important part of the process of software
development is the testing phase. According to good practices, this should
always include a suite of tests devoted to the analysis of the performance.
Various tools exist to assist with the task, which in general depend on the
platform and the run-time. Benchmark figures are generally the main performance
estimators, which can also give an indication of whether a performance
regression has occurred during the development stage. However, once regressions
have been identified, one turns to other tools, like \emph{profilers}, to try to
pin-point their exact location in the code.

Profilers can be classified into two main categories: \emph{deterministic} and
\emph{statistical}. An in-depth discussion about the difference between these
two classes is beyond the scope of this paper. Here we want to focus on how to
best make use of the data that these tools allow us to collect. During the
years, many ways of aggregating and presenting profiling data have been
conceived. Flame graphs are arguably the most popular visual way of representing
such data, which also highlights the hierarchical nature of profiling data,
which generally involves capturing branches of the call graph. The invention of
flame graphs is commonly attributed to Brendan Gregg \cite{gregg16}, and more
details can be found on their personal blog \cite{greggblog}.

\begin{figure}[t!]
    \centering
    \begin{tikzpicture}
\fill[fill=red!16] (0,0) rectangle (8,0.25) node[midway,scale=0.5] {Process 77905};
\fill[fill=red!46] (0,0.25) rectangle (8,0.5) node[midway,scale=0.5] {Thread 4428795392};
\fill[fill=red!23] (0,0.5) rectangle (0.07612,0.75) node[midway,scale=0.5] {};
\fill[fill=red!33] (0,0.75) rectangle (0.07612,1) node[midway,scale=0.5] {};
\fill[fill=red!33] (0,1) rectangle (0.07612,1.25) node[midway,scale=0.5] {};
\fill[fill=red!33] (0,1.25) rectangle (0.07612,1.5) node[midway,scale=0.5] {};
\fill[fill=red!21] (0,1.5) rectangle (0.07612,1.75) node[midway,scale=0.5] {};
\fill[fill=red!30] (0,1.75) rectangle (0.07612,2) node[midway,scale=0.5] {};
\fill[fill=red!16] (0,2) rectangle (0.07031,2.25) node[midway,scale=0.5] {};
\fill[fill=red!51] (0,2.25) rectangle (0.07031,2.5) node[midway,scale=0.5] {};
\fill[fill=red!18] (0,2.5) rectangle (0.07031,2.75) node[midway,scale=0.5] {};
\fill[fill=red!14] (0,2.75) rectangle (0.06484,3) node[midway,scale=0.5] {};
\fill[fill=red!41] (0,3) rectangle (0.06484,3.25) node[midway,scale=0.5] {};
\fill[fill=red!49] (0,3.25) rectangle (0.0541,3.5) node[midway,scale=0.5] {};
\fill[fill=red!50] (0,3.5) rectangle (0.0541,3.75) node[midway,scale=0.5] {};
\fill[fill=red!60] (0,3.75) rectangle (0.0541,4) node[midway,scale=0.5] {};
\fill[fill=red!30] (0,4) rectangle (0.0541,4.25) node[midway,scale=0.5] {};
\fill[fill=red!45] (0,4.25) rectangle (0.0541,4.5) node[midway,scale=0.5] {};
\fill[fill=red!38] (0,4.5) rectangle (0.0541,4.75) node[midway,scale=0.5] {};
\fill[fill=red!20] (0,4.75) rectangle (0.0541,5) node[midway,scale=0.5] {};
\fill[fill=red!44] (0,5) rectangle (0.0541,5.25) node[midway,scale=0.5] {};
\fill[fill=red!48] (0,5.25) rectangle (0.0541,5.5) node[midway,scale=0.5] {};
\fill[fill=red!37] (0,5.5) rectangle (0.0541,5.75) node[midway,scale=0.5] {};
\fill[fill=red!40] (0,5.75) rectangle (0.0541,6) node[midway,scale=0.5] {};
\fill[fill=red!14] (0,6) rectangle (0.0541,6.25) node[midway,scale=0.5] {};
\fill[fill=red!56] (0,6.25) rectangle (0.0541,6.5) node[midway,scale=0.5] {};
\fill[fill=red!66] (0,6.5) rectangle (0.03621,6.75) node[midway,scale=0.5] {};
\fill[fill=red!43] (0,6.75) rectangle (0.03621,7) node[midway,scale=0.5] {};
\fill[fill=red!11] (0,7) rectangle (0.03621,7.25) node[midway,scale=0.5] {};
\fill[fill=red!55] (0,7.25) rectangle (0.03621,7.5) node[midway,scale=0.5] {};
\fill[fill=red!16] (0,7.5) rectangle (0.03621,7.75) node[midway,scale=0.5] {};
\fill[fill=red!22] (0,7.75) rectangle (0.03621,8) node[midway,scale=0.5] {};
\fill[fill=red!63] (0,8) rectangle (0.01482,8.25) node[midway,scale=0.5] {};
\fill[fill=red!44] (0,8.25) rectangle (0.01482,8.5) node[midway,scale=0.5] {};
\fill[fill=red!21] (0,8.5) rectangle (0.01482,8.75) node[midway,scale=0.5] {};
\fill[fill=red!16] (0.0541,3.25) rectangle (0.06484,3.5) node[midway,scale=0.5] {};
\fill[fill=red!18] (0.0541,3.5) rectangle (0.06484,3.75) node[midway,scale=0.5] {};
\fill[fill=red!35] (0.07612,0.5) rectangle (0.09357,0.75) node[midway,scale=0.5] {};
\fill[fill=red!59] (0.07612,0.75) rectangle (0.09357,1) node[midway,scale=0.5] {};
\fill[fill=red!17] (0.07612,1) rectangle (0.08763,1.25) node[midway,scale=0.5] {};
\fill[fill=red!37] (0.09357,0.5) rectangle (7.73295,0.75) node[midway,scale=0.5] {[module]:21};
\fill[fill=red!17] (0.09357,0.75) rectangle (6.06297,1) node[midway,scale=0.5] {from\_austin:16};
\fill[fill=red!41] (0.09357,1) rectangle (0.16446,1.25) node[midway,scale=0.5] {};
\fill[fill=red!32] (0.16446,1) rectangle (0.39118,1.25) node[midway,scale=0.5] {};
\fill[fill=red!54] (0.39118,1) rectangle (4.43775,1.25) node[midway,scale=0.5] {\_\_add\_\_:5};
\fill[fill=red!52] (4.43775,1) rectangle (4.47472,1.25) node[midway,scale=0.5] {};
\fill[fill=red!59] (4.47472,1) rectangle (4.51568,1.25) node[midway,scale=0.5] {};
\fill[fill=red!24] (6.06297,0.75) rectangle (7.40921,1) node[midway,scale=0.5] {from\_austin:12};
\fill[fill=red!44] (6.06297,1) rectangle (6.14854,1.25) node[midway,scale=0.5] {};
\fill[fill=red!56] (6.14854,1) rectangle (6.16082,1.25) node[midway,scale=0.5] {};
\fill[fill=red!43] (6.16082,1) rectangle (6.17065,1.25) node[midway,scale=0.5] {};
\fill[fill=red!29] (6.17065,1) rectangle (6.29185,1.25) node[midway,scale=0.5] {};
\fill[fill=red!13] (7.40921,0.75) rectangle (7.70288,1) node[midway,scale=0.5] {};
\fill[fill=red!58] (7.40921,1) rectangle (7.44086,1.25) node[midway,scale=0.5] {};
\fill[fill=red!24] (7.70288,0.75) rectangle (7.73295,1) node[midway,scale=0.5] {};
\fill[fill=red!38] (7.73295,0.5) rectangle (7.74523,0.75) node[midway,scale=0.5] {};
\end{tikzpicture}
    \caption{An example of a flame graph. This particular instance has been
    obtained from a Python application using the high-performance frame stack
    sampler for CPython Austin \cite{austin},
    which provides line-level information.}
    \label{fig:fg}
\end{figure}
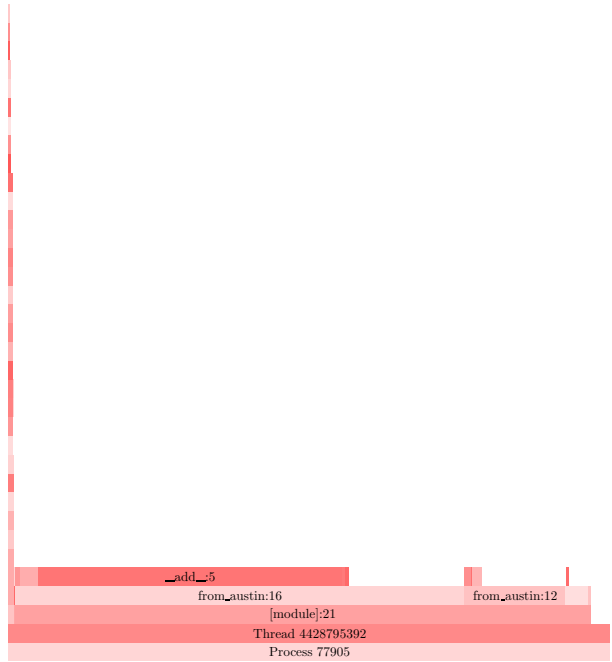

The aim of this paper is to propose a possible mathematical definition of what
flame graphs are. In doing so, we gain some interesting algebraic properties
almost for free, which in turn allow us to define some operations that can allow
to perform an in-depth performance regression analysis. The typical documented
use of a flame graph is via its graphical representation, whereby one scans the
picture for the largest plateaux. Whilst this method is effective at finding the
main sources of performance issues, it leaves quite a large amount of data
potentially unused. By combining a mathematical precise definition of flame
graphs with some statistical methods we show how to automate and enhance this
visual procedure and make the best of the full set of collected profiling data.

\section{Main Definitions}

The building blocks of a flame graph are the \emph{frames}. The information
available for each frame depends, in general, on the run-time, but here we shall
consider them as the fundamental unit on which flame graphs are built upon.

\begin{definition}[Frame set] Given the source code for a certain running
process, we shall denote by $F$ the set of all the possible frames that it can
generate.
\end{definition}
To make an analogy with the theory of formal languages, we can regard the set
$F$ as the \emph{alphabet} of a language.

Intuitively, a \emph{stack} of frames is a finite sequence of frames from a
frame set $F$. Formally, we can then think of a stack as an element of
$\bigcup_{n\in\mathbb{N}} F^n$, that is a tuple of a finite number $n$ of
elements from $F$, for some natural $n$. A stack is admissible by a certain set
of sources if it can occur at any point during the execution of the
corresponding program.
\begin{definition}[Stack set] The stack set of a certain set of sources is the
subset $S$ of $\bigcup_{n\in\mathbb{N}} F^n$ of all admissible stacks.
\end{definition}
Like the set of admissible words in a formal language is restricted by a set of
rules, similarly the set of all admissible stacks, say $S$, is restricted by the
admissible code paths. Hence, in general, the set $S$ will be a proper subset of
$\bigcup_{n\in\mathbb{N}} F^n$\footnote{In practice, the stack overflow problem
guarantees that we can always find $N\in\mathbb N$ such that
$S\subset\bigcup_{n=1,\ldots N} F^n$. Thus, we are allowed to make the
assumption that $S$ is a finite set in practical applications.}.

It should be clear from the previous discussion that the stack set $S$ depends
on the code base. However, it is conceivable that different code bases might
lead to the same stack set. This can happen, for example, if a function that
does not call into other functions is optimised in its computational complexity.
We shall therefore say that two code bases that lead to the same stack set are
\emph{code path-equivalent}. It might be worth considering equivalence relations
on the set of frames too, to avoid the issue of ending up with seemingly
inequivalent code bases when, e.g., functions are renamed. On the practical
level, one should then try to identify frames in a way that would yield a
reasonable notion of code path-equivalence. Defining equivalence relations on
the call-graph generated by a code base might help with this task.

\subsection{Flame Graphs}

Let $\Phi_S$ denote the free vector space generated by a stack set $S$ with
coefficients in $\mathbb{R}$. That is, $\Phi_S$ is the set of finite sums of
elements of $S$ multiplied by coefficients in $\mathbb R$ or, equivalently, as
the set of finitely supported functions over $S$ with values in $\mathbb R$.
Recall that the positive cone $\Phi_S^+$ of $\Phi_S$ is the set of all finite
sums of elements of $S$ with coefficients in $\mathbb R^+$.

\begin{definition}[Flame Graph] A flame graph is an element of the positive cone
$\Phi_S^+$ of $\Phi_S$.
\end{definition}

Based on the definition above, a possible way of describing a flame graph is by
giving a list of pairs, where each pair is made of a stack, that is a tuple of
frames, and a positive number. Indeed, this description coincides with what is
known as the \emph{collapsed} or \emph{folded} stack format \cite{gregg16}. In
this picture, the vector operation of adding two flame graphs together
corresponds to summing the value associated to each collapsed stack. This
aggregation procedure then yields a new flame graph, which corresponds to the
sum of the two original flame graphs.

\begin{figure}
    \centering
    \psset{xunit=1cm,yunit=.7cm}

\begin{pspicture}(8, 4)
    \psframe*[linecolor=red!30](0, 0)(8, 1)
    \psframe*[linecolor=red!20](0, 1)(1, 2)
    \psframe*[linecolor=red!50](1, 1)(6, 2)
    \psframe*[linecolor=red!40](1, 2)(5, 3)

    \psline[linestyle=dashed](1, 0)(1,3)
    \psline[linestyle=dashed](5, 0)(5,3)
    \psline[linestyle=dashed](6, 0)(6,3)

    \rput[b]{*0}(0.5, 2.25){1}
    \rput[b]{*0}(0.5, 1.25){B}
    \rput[b]{*0}(0.5, 0.25){A}

    \rput[b]{*0}(3, 3.25){5}
    \rput[b]{*0}(3, 2.25){D}
    \rput[b]{*0}(3, 1.25){C}
    \rput[b]{*0}(3, 0.25){A}

    \rput[b]{*0}(5.5, 2.25){1}
    \rput[b]{*0}(5.5, 1.25){C}
    \rput[b]{*0}(5.5, 0.25){A}

    \rput[b]{*0}(7, 1.25){2}
    \rput[b]{*0}(7, 0.25){A}
\end{pspicture}
    \caption{A simple flame graph. The vertical dashed lines separate the folded
    stacks which, in order from left to right, are: A;B, A;C;D, A;C and A. Their
    widths (the numbers on the top) encode the number that is associated with
    them. In this example, the vector representing this flame graph is
    $e_{\text{A;C}} + 5e_{\text{A;C;D}} + e_{C;A} + 2e_{A}$, where we use the
    notation $e_s$ to denote the basis vector of $\Phi_S$ associated with the
    stack $s\in S$.}
    \label{fig:folded_stacks}
\end{figure}
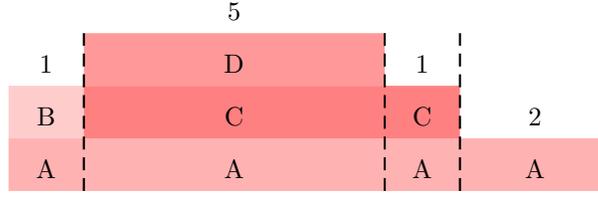

As it is probably known by the reader already, the name \emph{flame graph} is
derived by the way that folded stacks are normally represented. One starts by
grouping stacks based on the first frame in the stack tuple and laying them one
next to the other, with a width that equals the sum of all the coefficients in
each group. This process is repeated recursively within each group, by moving to
the next entry in the stack tuple (if any), and by drawing the frames on top of
each parent group. This yields a picture that resembles a free-burning flame,
whence the name. This visualisation based on frame grouping highlights further
structure that is perhaps not immediately visible in the folded format. Indeed,
in this visualisation one can create a \emph{hierarchy} of frames based on the
parent-child relation across all stacks, which is the reflection of the
caller-callee relation in code paths. This hierarchical structure, whilst very
interesting in its own right, will not be central in the discussions presented
in this paper, and might be explored further in future work.

\subsection{Flame Charts}

The flame graphs that one usually generates with the existing tools normally
come from the data collected by sampling statistical profilers. Deterministic
profiling data is usually the by-product of tools like tracers, which trace
function calls and emit events on entry and on exit. Contrary to sampled stacks,
this data is naturally suited to capture the hierarchical structure of profiles.
When the sequence of events is represented in the spirit of flame graphs, one
obtains a graph that is generally known as a \emph{flame chart} \cite{gregg16}.
In this case, the ``$x$-axis'' represents the passage of (wall) time, and the
visualisation is meant to represent the time at which the events occurred,
together with the parent-child relations that define the hierarchical structure.
In simple terms, one can then describe a flame chart as a time-ordered sequence
of flame graphs.

\begin{definition}[Flame Chart] A flame chart is a finite sequence $((t_1, f_1),
\ldots,(t_n, f_n))$ of elements from $\mathbb R\times\Phi_S^+$ with the property
that $t_i \leq t_j$ whenever $i < j$.
\end{definition}

Since $\Phi_S^+$ is closed under finite sums of its elements, we can see that we
can construct a flame graph from a given flame chart $\chi=((t_1, f_1),
\ldots,(t_n, f_n))$ by taking the sum of all of its flame graphs, viz.
    $$f_\chi = \sum_{i=1}^n f_n.$$
This aggregation process provides a link between deterministic and statistical
profiling, for we can regard the aggregation performed on statistical profiling
data as a way of providing a statistical representation of $f_\chi$. In some
visualisation tools, like \cite{speedscope}, this operation can be performed by
switching to the so-called \emph{left-heavy} mode, which is the process of
aggregating the same collapsed stacks to the left to produce a single flame
graph. Clearly, this procedure destroys the time information that is encoded in
the flame chart, with no possibility of recovering it.

\section{Flame Graph Algebra}

Once we have identified the flame graph space with the positive cone of a free
vector space, there are a few algebraic operations that can be performed on
flame graphs which essentially come for free. The aim of this section, however,
is to provide some order and attach a practical meaning to some of these
operations. Ultimately, the goal is to provide solid theoretical foundations to
the theory of flame graphs as useful tools for performance analysis of software.

The first observation that we make is that every element $\phi\in\Phi$ can be
decomposed in a special linear combination of two flame graphs, $\phi^+$ and
$\phi^-$, viz.
    $$\phi = \phi^+ - \phi^-.$$
Recall that by the definition of free vector space adopted in this paper we can
regard the elemets of $\Phi_S$ as finitely supported real functions over $S$. We
obtain $\phi^+$ from $\phi$ by simply setting to $0$ the function $\phi(s)$ over
$S$ wherever $\phi(s)$ is negative. We do the same with $-\phi$ to obtain
$\phi^-$, and it is easy to see that the previous identity is indeed correct,
and that $\phi^\pm$ are both flame graphs.

By the very definition, when we regard a flame graph as a map $f: S\to\mathbb
R$, the support of $f$ is always \emph{finite}. This checks out with the flame
graphs that we see in reality, as infinite flame graphs would require an
infinite amount of storage and an infinite observation time, which are not
features that can be achieved with physical computers.

\subsection{Differential Analysis}

The typical use of flame graphs as performance analysis tools relies quite
heavily on their visualisation. Standard techniques are based on the process of
collecting profiling data in the form of folded stacks, and visualise them in
the form of a flame graph, as described earlier. One then normally looks at the
largest plateaux occurring in the visualisation, which are normally associated
with the heaviest computational units. Whilst proven powerful in practice, this
method arguably lacks a systematic approach, and doesn't leverage the full data
set.

One task that might be difficult to perform when working with a visual
representation of flame graphs is \emph{comparison}. A method was proposed by
Gregg in \cite{gregg16,greggdiff}, with their idea of \emph{red-blue}
differential flame graphs. The procedure to generate these graphs is as follows.
One starts by drawing the flame graph obtained after a code change. Common
frames are then colored \emph{blue} if their value in the first profile is
higher, and \emph{red} otherwise.

With the algebra that we were able to endow on flame graphs, we can make the
idea of a difference between flame graphs more rigorous from a mathematical
point of view. Suppose that we have two flame graphs, $f_1$ and $f_2$. Their
difference $\Delta$, say
    $$\Delta := f_2 - f_1,$$
is not a flame graph in general, since the positive cone is not closed under
subtraction, but will certainly be an element of the free vector space $\Phi_S$.
Now we have already argued that any element of $\Phi_S$ can be decomposed into
the pair of two flame graphs. So, in this case, we know that we can find two
flame graphs, $\Delta^\pm$, such that
\begin{equation}\label{eq:delta}
    \Delta = \Delta^+ - \Delta^-.
\end{equation}

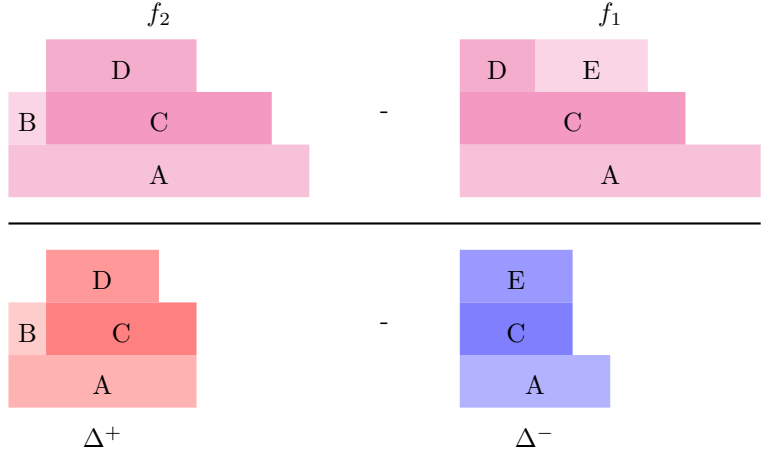
\begin{figure}
    \centering
    \psset{xunit=0.5cm,yunit=.7cm}

\begin{pspicture}(0,-1)(20, 8)
    \rput[b]{*0}(4, 7.25){$f_2$}

        \psframe*[linecolor=magenta!30](0, 4)(8, 5) 
        \psframe*[linecolor=magenta!20](0, 5)(1, 6) 
        \psframe*[linecolor=magenta!50](1, 5)(7, 6) 
        \psframe*[linecolor=magenta!40](1, 6)(5, 7) 

        \rput[b]{*0}(0.5, 5.25){B}
        \rput[b]{*0}(3, 6.25){D}
        \rput[b]{*0}(4, 5.25){C}
        \rput[b]{*0}(4, 4.25){A}

    \rput[b]{*0}(10, 5.5){-}

    \rput[b]{*0}(16, 7.25){$f_1$}

        \psframe*[linecolor=magenta!30](12, 4)(20, 5) 
        \psframe*[linecolor=magenta!50](12, 5)(18, 6) 
        \psframe*[linecolor=magenta!40](12, 6)(14, 7) 
        \psframe*[linecolor=magenta!20](14, 6)(17, 7) 

        \rput[b]{*0}(15.5, 6.25){E}
        \rput[b]{*0}(13, 6.25){D}
        \rput[b]{*0}(15, 5.25){C}
        \rput[b]{*0}(16, 4.25){A}

    \psline(0, 3.5)(20, 3.5) 

    \rput[b]{*0}(2.5, -0.75){$\Delta^+$}

        \psframe*[linecolor=red!30](0, 0)(5, 1) 
        \psframe*[linecolor=red!20](0, 1)(1, 2) 
        \psframe*[linecolor=red!50](1, 1)(5, 2) 
        \psframe*[linecolor=red!40](1, 2)(4, 3) 

        \rput[b]{*0}(0.5, 1.25){B}
        \rput[b]{*0}(2.5, 2.25){D}
        \rput[b]{*0}(3, 1.25){C}
        \rput[b]{*0}(2.5, 0.25){A}

    \rput[b]{*0}(10, 1.5){-}

    \rput[b]{*0}(14, -0.75){$\Delta^-$}

        \psframe*[linecolor=blue!30](12, 0)(16, 1) 
        \psframe*[linecolor=blue!50](12, 1)(15, 2) 
        \psframe*[linecolor=blue!40](12, 2)(15, 3) 

        \rput[b]{*0}(13.5, 2.25){E}
        \rput[b]{*0}(13.5, 1.25){C}
        \rput[b]{*0}(14, 0.25){A}

\end{pspicture}
    \caption{Visual representation of the difference between two flame graphs.
    In this example, the $\Delta^+$ part shows that there is an increase in the
    value for D and \emph{own} value, that is the part that is not covered by
    child frames, for C, as well as the appearance of B. The $\Delta^-$ part
    shows a disappearance of the E stack, as well as a decrease in own value
    for A.}
    \label{fig:red_blue}
\end{figure}

What meaning can we attach to $\Delta^\pm$? Let $\sigma_1$ and $\sigma_2$ be the
support of $f_1$ and $f_2$ respectively, and let $\sigma_i = \sigma_1 \cap
\sigma_2$. Clearly, $\Delta^+(s) = f_2(s)$ for any $s \in
\sigma_2\smallsetminus\sigma_i$, and like-wise $\Delta^-(s) = f_1(s)$ for any
$s\in\sigma_1\smallsetminus\sigma_i$. That is, $\Delta^+$ is the same as $f_2$
on those stacks in the support of $f_2$ that are not common to the support of
$f_1$ (and similarly for $\Delta^-$ and $f_1$). So the more interesting
situation is the description of $\Delta^\pm$ on the support intersection
$\sigma_i=\sigma_1\cap\sigma_2$. We can see that $\Delta^+(s)$ equals the
difference $f_2(s)-f_1(s)$ on every stack $s\in\sigma_i$ where $f_2(s)>f_1(s)$,
and that $\Delta^-(s) = f_1(s) - f_2(s)$ on every $s\in\sigma_i$ where $f_1(s) >
f_2(s)$.

Suppose now that $f_1$ and $f_2$ are the CPU profiles of two code
path-equivalent code bases. We can see that $\Delta^+$ ($\Delta^-$) carries
information about new code paths not exercised when $f_1$ ($f_2$) was collected,
plus the additional CPU times in $f_2$ ($f_1$) of stacks that are common to
$f_1$ ($f_2$). If $\supp f_1$ = $\supp f_2$, then $\Delta^+$ ($\Delta^-$)
highlights the stacks that used more (own) CPU when running the second (first)
code base, compared to the same stack in the first (second) profile.

What happens if we compare flame graphs generated from code bases that are not
code path-equivalent? Nobody stops us from considering the union of the frames
of each code base, and hence the admissible stacks for both of them, say $S\cup
S'$, and then embed the flame graphs in $\Phi_{S\cup S'}$. However, the analysis
we have described in the previous paragraph becomes harder, as we might have
different code paths that perhaps perform the same computations, but that are
not so immediately relatable. In this case, more traditional benchmark
techniques might be more suited for performance analysis and regression
detection.

Before moving on to the next section, we would like to provide yet another way
of looking at the decomposition \eqref{eq:delta}. Indeed, we can write
    $$\Delta = (\Delta^+_a + \Delta^+_g) - (\Delta^-_d + \Delta^-_s)$$
where all the terms on the RHS are flame graphs with mutually disjoint supports.
The term $\Delta^+_a$ has a support with empty intersection with the support of
$f_1$, and we interpret it as the flame graph of stacks that have appeared in
$f_2$. Similarly $\Delta^-_d$ represent the flame graphs of stacks in $f_1$ that
have \emph{disappeared} in $f_2$. The term $\Delta^+_g$ represents the stacks
common to $f_1$ and $f_2$ but that have \emph{grown} from $f_1$ to $f_2$ (i.e.
have positive delta); similarly, $\Delta^-_s$ represents those stacks that have
shrunk in the passage from $f_1$ to $f_2$. This kind of decomposition could be
useful when one needs to validate extra assumptions about code paths. For
example, if one is trying to determine the overhead profile of a tracer, one
would not expect a non-trivial term $\Delta^+_a$ (after having filtered out any
frames contributed  by the tracer itself), as that might be an indication that
the tracer is somehow changing code paths and interacting in an expected way
with the tracee. As for the overhead itself, the component $\Delta^+_g$ should
provide the sought profile in this case. The presence of non-trivial $\Delta^-$
terms might provide some further insight too. Shrinking stacks might be the
indication that the system is under saturation, and therefore some resources
have been shifted to other code paths, as a result of the operation of the
tracer on the tracee.

\begin{figure}
    \centering
    \psset{xunit=0.5cm,yunit=.7cm}

\begin{pspicture}(0,-1)(20, 8)
    \rput[b]{*0}(4, 7.25){$f_2$}

        \psframe*[linecolor=magenta!30](0, 4)(8, 5) 
        \psframe*[linecolor=magenta!20](0, 5)(1, 6) 
        \psframe*[linecolor=magenta!50](1, 5)(7, 6) 
        \psframe*[linecolor=magenta!40](1, 6)(5, 7) 

        \rput[b]{*0}(0.5, 5.25){B}
        \rput[b]{*0}(3, 6.25){D}
        \rput[b]{*0}(4, 5.25){C}
        \rput[b]{*0}(4, 4.25){A}

    \rput[b]{*0}(10, 5.5){-}

    \rput[b]{*0}(16, 7.25){$f_1$}

        \psframe*[linecolor=magenta!30](12, 4)(20, 5) 
        \psframe*[linecolor=magenta!50](12, 5)(18, 6) 
        \psframe*[linecolor=magenta!40](12, 6)(14, 7) 
        \psframe*[linecolor=magenta!20](14, 6)(17, 7) 

        \rput[b]{*0}(15.5, 6.25){E}
        \rput[b]{*0}(13, 6.25){D}
        \rput[b]{*0}(15, 5.25){C}
        \rput[b]{*0}(16, 4.25){A}

    \psline(0, 3.5)(20, 3.5) 

    \rput[b]{*0}(0.5, -0.75){$\Delta^+_a$}
    \rput[b]{*0}(4, -0.75){$\Delta^+_g$}

        \psframe*[linecolor=red!30](0, 0)(1, 1) 
        \psframe*[linecolor=red!20](0, 1)(1, 2) 

        \psframe*[linecolor=red!30](2, 0)(6, 1) 
        \psframe*[linecolor=red!50](2, 1)(6, 2) 
        \psframe*[linecolor=red!40](2, 2)(5, 3) 

        \rput[b]{*0}(0.5, 1.25){B}
        \rput[b]{*0}(0.5, 0.25){A}

        \rput[b]{*0}(3.5, 2.25){D}
        \rput[b]{*0}(4, 1.25){C}
        \rput[b]{*0}(4, 0.25){A}

    \rput[b]{*0}(10, 1.5){-}

    \rput[b]{*0}(13.5, -0.75){$\Delta^-_d$}
    \rput[b]{*0}(16.5, -0.75){$\Delta^-_s$}

        \psframe*[linecolor=blue!30](12, 0)(15, 1) 
        \psframe*[linecolor=blue!50](12, 1)(15, 2) 
        \psframe*[linecolor=blue!40](12, 2)(15, 3) 

        \psframe*[linecolor=blue!30](16, 0)(17, 1) 

        \rput[b]{*0}(13.5, 2.25){E}
        \rput[b]{*0}(13.5, 1.25){C}
        \rput[b]{*0}(13.5, 0.25){A}

        \rput[b]{*0}(16.5, 0.25){A}

\end{pspicture}
    \caption{Explicit visualisation of all the $\Delta$ terms from the example
    of Figure~\ref{fig:red_blue}.}
    \label{fig:red_blue_2}
\end{figure}
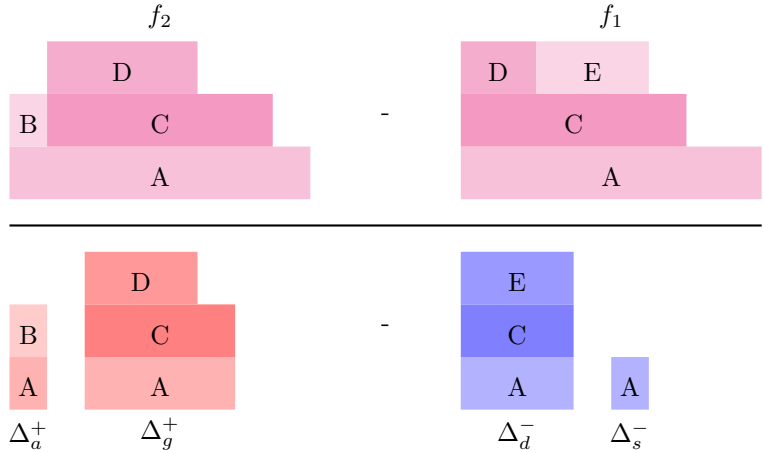

Finally, we note that, when it comes to visualising the $\Delta$ flame graphs,
it might be wise to normalise by the norm of $f_1$ or $f_2$, as appropriate.
This is because the information of how big of a difference one has gets lost in
the process. This normalisation step can then provide a relative measure of how
much the stacks have grown/shrunk etc\ldots in relative terms.

\subsection{Metrisation}

Most often than not, flame graphs are generated by \emph{sampling} profilers,
which are statistical in nature. Running the same code, under the same
conditions, through the same profiler, is bound to generate different flame
graphs, albeit with a very similar shape. But how can we \emph{quantify} the
resemblance of two flame graphs?

From our formalisation of the notion of flame graphs carried out in the previous
section, we have seen that we can regard a flame graph as an element of (the
positive code of) a vector space, viz. $\Phi_S$. Vector spaces can be
\emph{normed}, that is they can be equipped with a notion of \emph{length} of
their elements, which can then be used to define a distance function that gives
a measure of how different two elements are. For the elements of $\Phi_S$ we
could choose the norm
    $$\Vert\phi\Vert = \sum_{s\in S}|\phi(s)|$$
and the distance function
    $$d(\phi,\psi) = \Vert\phi-\psi\Vert.$$
Practically, if $f$ is the flame graph of a CPU profile, then $\Vert f\Vert$
measures the total CPU time that has been recorded.

We observe that
    $$\Vert\phi-\psi\Vert \leq\Vert\phi\Vert + \Vert\psi\Vert,$$
for any $\phi,\psi\in\Phi_S$, with the equality attained whenever $\phi$ and
$\psi$ have disjoint support. Therefore, a na\"ive similarity measure between
flame graphs can be defined as
    $$\sigma(f,g) = 1 - \frac{\Vert f - g\Vert}{\Vert f\Vert + \Vert g\Vert}.$$
The closer this score is to 1, the more ``similar'' the flame graphs are.

\subsection{Performance Regression Detection}

If the source of the flame graphs is deterministic in nature (e.g. as the
outcome of running the code base through a deterministic profiler), the
similarity score of the previous section can give an idea of how far apart two
such flame graphs are. A value different from 1 would then signal an immediate
performance profile change and prompt an in-depth performance analysis.

When the flame graphs originate from a sampling process, the similarity score
$\sigma$ is of no much use and it is perhaps best to consider conducting a
statistical analysis of the collected data. For instance, if the goal is to
detect potential performance regression in code path-equivalent code bases, one
possibility is to collect a statistically significant sample of flame graphs.
When running performance tests, a new sample of flame graphs can be collected,
and their distribution tested against the original one, e.g. via a two-sample
Hotelling $T^2$ test.

This kind of analysis allows to answer the question of which stacks are
responsible for a failed hypothesis test too, in a way that generalises the more
visual approach of spotting the largest plateaux, e.g. in a differential flame
graph, or across two such graphs laid side-by-side. Indeed, with a critical
$p$-value of $p^*$, say $1\%$, one would find the critical value $F^*$ for the
$F$-distributed statistic\footnote{via the inverse cumulative distribution
function}
    $$f = G^2\Delta^T \Sigma_p^{-1}\Delta\ \sim\ F(p, n_1 + n_2 - p - 1),$$
where $G^2$ is the constant
    $$G^2=\frac{n_1 + n_2 - p - 1}{(n_1 + n_2 - 2)p}\frac{n_1n_2}{n_1+n_2},$$
$n_1$ and $n_2$ are the number of points in each of the two samples
respectively, $\Delta$ is the difference between the average flame graphs from
each sample, $p$ is the cardinality of the union of the supports of the average
flame graphs, and $\Sigma_p$ is the \emph{pooled} covariance matrix. We can find
the simultaneous $F^*$-confidence intervals for each stack $s_k$ by solving the
constraint problem
\begin{equation*}
    \begin{cases}
        G^2(s-\Delta)^T\Sigma_p^{-1}(s-\Delta) = F^*\\
        \nabla_s G^2(s-\Delta)^T\Sigma_p^{-1}(s-\Delta) = \pm\lambda e_k
    \end{cases}
\end{equation*}
where $\lambda > 0$ is a Lagrange multiplier and $e_k$ is the basis vector in
the direction of the stack variable $s_k$. We find
    $$s - \Delta = \pm\lambda \Sigma_p e_k$$
which substituted into the first equation yields
    $$\lambda = \sqrt{\frac{F^*}{G^2 \diag(\Sigma_p)_k}}$$
where $\diag(\Sigma_p)_k$ is the $k$-th diagonal element of $\Sigma_p$, i.e. the
pooled variance of the $k$-th stack, viz $\diag(\Sigma_p)_k = (e_k, \Sigma_p
e_k)$. The confidence interval for the $k$-th stack is then given by
    $$\left[\Delta_k - \sqrt{\frac{F^*\diag(\Sigma_p)_k}{G^2}}, \Delta_k + \sqrt{\frac{F^*\diag(\Sigma_p)_k}{G^2}} \right].$$
The stacks responsible for a failed hypothesis test are those for which the
confidence interval does not contain the expected null difference between the
means, i.e. those for which one has
    $$\Delta_k^2 > \frac{F^*\diag(\Sigma_p)_k}{G^2}.$$

\begin{example} Suppose that we collected 100 flame graphs for a code base, made
a code change somewhere, and collected other 100 flame graphs with the new code
to check for any potential performance regressions. Let us further assume that
the average profiles show just three stacks, A, B and C, with the following
metrics
$$f_1 = [1e5, 2e5, 3e5]\qquad f_2 = [1.001e5, 4e5, 2.998e5]$$
and the pooled covariance matrix has diagonal entries given by 5000, 7500 and
10000 respectively. With the method described above, and a critical $F$-value of
$F^*=3.8$\footnote{which roughly corresponds to a $p$-value of $1\%$}, we find
the confidence intervals
$$[-140, 340],\quad [199700, 200300],\quad [-539, 139].$$
This immediately indicates that there is a clear difference between the two
profiles, and that B is the statistically significant different stack between
them. If the metric attached to each stack in this example is CPU time, this
result indicates that stack B is taking more CPU time after the code change,
which could indicate a potential performance regression along the code path
represented by the stack B.
\end{example}

\begin{remark} The Hotelling $T^2$ test can be applied to normally distributed
multivariate samples. It is therefore important that experiments are carefully
designed to avoid any of the standard issues that arise when benchmarking, e.g.
periodic phenomena, unforseen correlations, noisy neighbours, thermal
throttling, to name a few. A comprehensive survey of issues and challenges in
benchmarking can be found in \cite{bartz2020benchmarking}. In general, one might
be \emph{rescued} by the Central Limit Theorem, but it is important to be aware
of the potential pitfalls.    
\end{remark}

On the practical side, collected flame graphs tend to have high dimensionality
(that is, we should consider vector spaces of pretty large dimensions, generally
order of 100s, or more). The problem with conducting a two-sample Hotelling test
on such high-dimensional data is to ensure that each sample has more
measurements than distinct frame stacks. One technique that could be adopted to
reduce the dimensionality to something more tractable is perhaps to construct a
frequency table of the frame stacks, with the goal of retaining only those that
appear more frequently with the sample, that is, a procedure akin to document
frequency in text document analysis. We expect this procedure to drop only frame
stacks that could be considered \emph{spurious} for the purposes of the
performance analysis.

\subsection{Differential Analysis Revisited}

The differential analysis of frame graphs presented previously makes the most
sense when applied to flame graphs that arise from a deterministic process. When
we are in a sampling setting, the same considerations around the similarity
score apply to the difference between flame graphs. Even when the differential
flame graph $\Delta$ arises as a difference between two \emph{average} flame
graphs, the result could still contain a fair amount of what we could call
\emph{statistical noise}\footnote{Note that we are not considering questions
about the accuracy of the data collected by a statistical profiler here; in
fact, we shall assume that, in this regard, the accuracy of the collected data
is to the satisfaction of the user. On the topic of sampling profilers and
accuracy we refer the reader to existing literature, e.g. \cite{embedded}}. One
way to reduce it is to use the confidence interval analysis that we carried out
at the end of the previous section. For example, once we have determined the set
$S_s$ of stacks that are responsible for the statistically significant
difference between two flame graphs, we can \emph{reduce} the support of
$\Delta$ by setting to 0 all stacks in $\supp\Delta$ that do not belong to
$S_s$, to obtain a new difference $\Delta_r$. This way we have removed those
differences that we cannot distinguish from 0 from a statistical point of view.
We can then use $\Delta_r$ to operate the decomposition that we have described
previously to carry out a more in-depth analysis with less statistical noise.

\section{Experimental results}

In this section we present the results of a simple experiment that shows the
method of the statistical differential analysis of the previous section. The
scenario we mimic is that of a team that has decided to start doing performance
analysis of their code changes. A performance regression has been identified and
a fix has been proposed. Before it is merged into the main branch, the team
want to make sure that the performance issue (which we mimic with a sleep) has
actually been resolved. They then collect 50 flame graphs for the new code and
50 flame graphs for the last released version. In carrying out the analysis
discussed in the previous section they also observe an unexpected $\Delta^+$
component, coming for a custom \texttt{sitecustomize.py} script that is executed
at startup, likely due to a previous, yet unreleased, change.

The code that we use to mimic the scenario we just described consists of a
\texttt{main.py} script that simulates the main application, which reads
\begin{lstlisting}[language=python]
    import os
    from time import sleep
    
    
    def a():
        sleep(0.15 if os.getenv("REGRESSION", False) else 0.2)
    
    
    def b():
        a()
        sleep(0.1)
    
    
    def c():
        b()
        sleep(0.05)
    
    
    c()
\end{lstlisting}    
and a \texttt{sitecustomize.py} script for the start-up initialisation, which
reads
\begin{lstlisting}[language=python]
    import os
    import time

    if os.getenv("REGRESSION", False):
        time.sleep(0.1)
\end{lstlisting}    

We can toggle the two different performance behaviours with the
\texttt{REGRESSION} environment variable. The baseline behaviour is that of the
previous release, which we obtain by not setting the environment variable.
Running the differential analysis of the previous section on the data we
collected with Austin 3.4.1 and using CPython 3.9.16, we obtain a $\Delta^-$
component with the c;b;a call stack, and a delta value of 50 ms. The unexpected
$\Delta^+$ component has the single stack of the \texttt{sitecustomize.py}
execution by the CPython module loader, with a delta of 100 ms. These values are
in perfect agreement with the pauses introduced in the two scripts\footnote{The
code for the experiment, and the instructions about how to run it, can be found
at \url{https://github.com/P403n1x87/flamegraph-experiment}}.

\section{Acknowledgements}

The author would like to thank Michał J Gajda for their valuable feedback and
suggestions on an earlier draft of this paper.


\bibliography{refs}{}\bibliographystyle{plain}

\begin{thebibliography}{1}

\bibitem{bartz2020benchmarking}
Thomas Bartz-Beielstein, Carola Doerr, Daan van~den Berg, Jakob Bossek, Sowmya
  Chandrasekaran, Tome Eftimov, Andreas Fischbach, Pascal Kerschke, William
  La~Cava, Manuel Lopez-Ibanez, et~al.
\newblock {Benchmarking in optimization: Best practice and open issues}.
\newblock {\em arXiv preprint arXiv:2007.03488}, 2020.

\bibitem{greggblog}
Brendan Gregg.
\newblock {Flame Graphs}.
\newblock \url{https://www.brendangregg.com/flamegraphs.html}.

\bibitem{greggdiff}
Brendan Gregg.
\newblock {Differential Flame Graphs}.
\newblock
  \url{https://www.brendangregg.com/blog/2014-11-09/differential-flame-graphs.html},
  2014.

\bibitem{gregg16}
Brendan Gregg.
\newblock The flame graph: This visualization of software execution is a new
  necessity for performance profiling and debugging.
\newblock {\em Queue}, 14(2):91–110, mar 2016.

\bibitem{embedded}
Embdedded Staff.
\newblock {Statistical Profiling: An Analysis}.
\newblock \url{https://www.embedded.com/statistical-profiling-an-analysis/},
  2000.

\bibitem{austin}
Gabriele~N. Tornetta.
\newblock {Austin: A Frame Stack Sampler for CPython}.
\newblock \url{https://github.com/P403n1x87/austin}, 2018.

\bibitem{speedscope}
Jamie Wong.
\newblock {Speedscope: A fast, interactive web-based viewer for performance
  profiles}.
\newblock \url{https://www.speedscope.app/}, 2017.

\end{thebibliography}

\end{document}